\documentclass[11pt]{article}
\usepackage[latin9]{inputenc}
\usepackage{geometry}
\usepackage{amsmath}
\usepackage{amssymb}
\usepackage{amsthm}
\usepackage{algorithm}
\usepackage{algorithmic}
\usepackage{float}
\usepackage{graphicx}
\usepackage{mathtools}
\usepackage{pgffor}
\usepackage[numbers, sort]{natbib}
\usepackage{fancyhdr}
\usepackage[title,titletoc]{appendix}
\usepackage{hyperref}

\graphicspath{{figures/}}
\makeatletter

\usepackage{amsfonts}\usepackage{nopageno}
\newcommand{\diag}{{\rm diag}}
\reversemarginpar

\newcommand{\be}{\begin{equation}}
\newcommand{\ee}{\end{equation}}
\newcommand{\mR}{{\mathbb R}}

\makeatother

\begin{document}
\title{Regularized Optimal Mass Transport with Nonlinear 
Diffusion}
\author{Kaiming Xu, Xinan Chen, Helene Benveniste, Allen Tannenbaum 
\thanks{K.\ Xu is with the Department of Applied Mathematics \& Statistics, Stony Brook University, NY; email: \href{mailto:kaiming.xu@stonybrook.edu}{kaiming.xu@stonybrook.edu}}
\thanks{X.\ Chen is with the Department of Medical Physics, Memorial Sloan Kettering Cancer Center, NY}
\thanks{H.\ Benveniste is with the Department of Anesthesiology, Yale School of Medicine, CT}
\thanks{A.\ Tannenbaum is with the Departments of Computer Science and Applied Mathematics \& Statistics, Stony Brook University, NY; email: \href{mailto:allen.tannenbaum@stonybrook.edu}{allen.tannenbaum@stonybrook.edu}}
}
\date{\today}
\maketitle
\thispagestyle{fancy}
      \lhead{}
      \chead{}
      \rhead{}
      \lfoot{}
      \rfoot{}
      \cfoot{\thepage}
      \renewcommand{\headrulewidth}{0pt}
      \renewcommand{\footrulewidth}{0pt}
\pagestyle{fancy}
\cfoot{\thepage}
\begin{abstract}
    In this paper, we combine nonlinear diffusion with the regularized optimal mass transport (rOMT) model. As we will demonstrate, this new approach provides further insights into certain applications of fluid flow analysis in the brain. From the point of view of image processing, the  anisotropic diffusion method, based on Perona-Malik, explicitly considers edge information. Applied to rOMT analysis of glymphatic transport based on dynamic 
    contrast-enhanced magnetic resonance imaging data, this new framework appears to capture a  larger advection-dominant volume. 
\end{abstract}

\section{Introduction}
The theory of optimal mass transport(OMT) was first proposed by Gaspard Monge in 1781 and has since evolved into a unique scientific field which has had significant impact on research in many disciplines  \cite{villani2021topics,villani2009optimal}.  Mass transport theory has been applied to diverse fields including physics, biology, economics and engineering. OMT defines a distance called the {\em Wasserstein distance},  and thus creates a natural geometry on the space of probability distributions. Our study is based on a fluid dynamics reformulation of OMT  \cite{benamou2000computational} which allows us to calculate the flow fields between two density distributions. 

Regularized optimal mass transport (rOMT), an extension of fluid dynamics reformulation of OMT, is a tool to study temporal flow fields as a physically inspired model of optical flow. It has the ability to capture the flow dynamics, handle noise and simulate diffusion \cite{elkin2018glymphvis,koundal2020optimal,chen2022cerebral}. rOMT utilizes an advection-diffusion equation as its flow-driven partial different equation and is endpoint free. A source term may be added to rOMT in which case the total mass preservation condition can be circumvented. This line of research will be pursued in other work. 

Anisotropic diffusion, a major tool for image segmentation, edge detection and image denoising, was first proposed by Perona and Malik \cite{perona1990scale}. Notably, instead of using a constant diffusion coefficient, Perona and Malik considered a nonnegative function ({\bf conductivity coefficient}) of the magnitude of the local density gradient; see equation~(\ref{anisotropicdiffusion}). The authors suggested two possible conductivity coefficients (see (\ref{sigma1}) and (\ref{sigma2})), wherein the diffusion will be very small near the edges, i.e. reflecting the fact that near edges images tend to have very large intensity gradients. In this work, we show  that anisotropic diffusion enhances the interpretation of glymphatic dynamic 
contrast-enhanced magnetic resonance imaging (DCE-MRI) flow data and may be used in conjunction with the constant diffusion coefficient approach  \cite{chen2022cerebral}. The anisotropic diffusion equation may be derived via the steepest descend method for solving an energy minimization problem \cite{you1996behavioral}.

The glymphatic system is involved in transporting waste products from the brain to the meningeal lymphatic system which connects to the cervical lymph nodes  \cite{nedergaard2013garbage}. The functioning of the glymphatic and lymphatic systems decrease with age and have been implicated in the pathophysiology of a wide range of neurodegenerative diseases including cerebral amyloid angiopathy \cite{chen2022cerebral,xu2022glymphatic} and Alzheimer's disease \cite{kress2014impairment,ma2017outflow,da2018functional,peng2016suppression}. We study glymphatic transport using a temporal series of DCE-MRI data acquired from the rodent brain \cite{iliff2013brain,lee2018quantitative,lee2015effect}. Since the data are acquired at discrete time points, our work is motivated by the need to find a dynamic physically based model of the transport. Several different versions of OMT \cite{ratner2017cerebrospinal} and rOMT \cite{elkin2018glymphvis,koundal2020optimal,chen2022cerebral} have been used to model the glymphatic flow.

In the present work, we propose a new version of rOMT. Specifically, we replace the linear diffusion in rOMT \cite{elkin2018glymphvis,koundal2020optimal,chen2022cerebral} with the Perona-Malik based anisotropic diffusion. Here, we argue that this gives us enhanced flexibility to study image-based flows inherent to glymphatic transport. Notably, many diffusion processes in fluids are better captured by nonlinear models, e.g., axisymmetric surface diffusion \cite{bernoff1998axisymmetric} and thin fluid films  \cite{king2001two,king2000emerging}. 
We utilize Lagrangian coordinates for visualizing 
the glymphatic transport pathlines.
Several properties of solute particle movement are computed along the pathlines such as speed and the P\'{e}clet number. Here we compare various parameters of the anisotropic diffusion coefficient, and observe the impact of different values on several data metrics including P\'{e}clet plots which can map diffusion dominated versus advection dominated regions of the brain.

We briefly summarize the contents of the present paper. 
In Section~\ref{model}, we review the theory of OMT, rOMT and nonlinear diffusion. Section~\ref{numeric} introduces the algorithm and numerical methods we employ for our current work. In Section~\ref{results}, we explicate the application of the model to glymphatic DCE-MRI data and analyze the experimental results and we conclude our paper in Section~\ref{discussion}.

\section{Model}\label{model}
\subsection{OMT}
In this section, we introduce OMT and its fluid dynamics formulation. All the technical details as well as a complete set of references may be found in \cite{villani2021topics,villani2009optimal}.
The original formulation of OMT was given by  Gaspard Monge and may be expressed as
\begin{equation}\label{originalOMT}
    \inf_T\{\int_\Omega c(x,T(x))\rho_0(x)dx \ | \ T_{\#}\rho_0=\rho_1\},
\end{equation}
where $c(x,y)$ is the cost function of moving the unit mass from $x$ to $y$, $\rho_0$ and $\rho_1$ are two probability distributions in the domain $\Omega\subseteq\mR^d$, $T$ is the transport map, and $T_{\#}$ is the push-forward of $T$. This formulation assumes that  $\rho_0$ and $\rho_1$ have the same total mass, i.e. $\int_\Omega\rho_0(x)dx=\int_\Omega\rho_1(x)dx$ and then seeks for the optimal transport map $T$ to minimize the total cost, the integral in equation (\ref{originalOMT}), subject to the push-forward constraint.

Later, Leonid Kantorovich formulated a relaxed version of OMT as follows:
\begin{equation}\label{relaxedOMT}
    \inf_{\pi\in\Pi(\rho_0,\rho_1)}\int_{\Omega\times \Omega}c(x,y)\pi(dx,dy),
\end{equation}
where $\Pi(\rho_0,\rho_1)$ denotes the set of all couplings (joint distributions) between the marginals $\rho_0$ and $\rho_1$. From here on, the cost function $c$ will be taken as the square of the Euclidean distance $c(x,y) = \|x-y\|^2.$

Benemou and Brenier \cite{benamou2000computational} proved that for $c(x,y)=\|x-y\|^2$, the specific infimum of Monge-Kantorovich formulation is equal to the result in following fluid dynamics formulation for density/probability distributions with compact support:
\begin{align}
    \inf_{\rho,v} &\int_0^1\int_{\Omega} \rho(t,x)|v(t,x)|^2dxdt, \label{omtenergyfunc}\\
    &\frac{\partial\rho}{\partial t}+\nabla\cdot(\rho v) = 0, \label{ae}\\
    &\rho(0,x) = \rho_0(x), \quad \rho(1,x) = \rho_1(x),  \label{marginal}
\end{align}
where $\rho:[0,1]\times\Omega\to\mR_{\ge0}$ is the family of density/probability distributions defining geodesic path from $\rho_0$ to $\rho_1$, and $v:[0,1]\times\Omega\to\mR^d$ is the velocity vector field.

\subsection{rOMT}

The regularized OMT model (rOMT) \cite{elkin2018glymphvis,koundal2020optimal} adds two assumptions: 1. the image data we use are noisy observations and thus we do not want to make the final density we calculate coincide with the MR images; and 2. the flow is driven by an advection-diffusion equation. Based on these two assumptions, the rOMT formulation may be written as: 
\begin{align}
    \inf_{\rho,v} &\int_0^1\int_{\Omega} \rho(t,x)|v(t,x)|^2dxdt+\beta\int_{\Omega}(\rho(1,x)-\rho_1(x))^2dx, \label{romtenergyfunc}\\
    &\frac{\partial\rho}{\partial t}+\nabla\cdot(\rho v) = \nabla\cdot(\sigma_0\nabla\rho), \label{ade}\\
    &\rho(0,x) = \rho_0(x). \nonumber
\end{align}
In this formulation, the final marginal condition is removed and a penalty of the error between final density and ground truth is added in the objective function (\ref{romtenergyfunc}), where $\beta$ is the penalty parameter. Equation (\ref{ade}) is an advection-diffusion equation with a constant $\sigma_0$ denoting the diffusion coefficient.

\subsection{Nonlinear diffusion}
Instead of using linear diffusion in which $\sigma_0$ is a constant, nonlinear diffusion seems to have certain advantages that we will now describe.  
Perona and Malik proposed an anisotropic diffusion \cite{perona1990scale}, which is a useful tool for image segmentation, edge detection and image denoising. The anisotropic diffusion equation is 
\begin{equation}\label{anisotropicdiffusion}
    \frac{\partial\rho}{\partial t} = \nabla\cdot(\sigma(|\nabla\rho|)\nabla\rho),
\end{equation}
where $\sigma(\cdot)$ is a nonnegative strictly decreasing function. If we consider a 3D problem, then $|\nabla\rho|=\sqrt{\rho_x^2+\rho_y^2+\rho_z^2}$. The proper diffusion should be large in smooth homogeneous areas and become smaller near edges, the places where $|\nabla\rho|$ is large. Perona and Malik \cite{perona1990scale} suggested two versions of the diffusion (conductivity) coefficient:
\begin{align}
    &\sigma(x) = \sigma_0\frac{1}{1+(\frac{x}{K})^2},\label{sigma1}\\
    &\sigma(x) = \sigma_0e^{-(\frac{x}{K})^2}.\label{sigma2}
\end{align}
Both are $0$ when $x$ approaches $\infty$ and attend upper bound $\sigma_0$ while $x=0$. $K$ is a constant and controls the sensitivity to edges and can be tuned for different applications.

Following \cite{you1996behavioral}, we may derive the anisotropic diffusion equation (\ref{anisotropicdiffusion}) via the steepest descent from an energy minimization problem. More precisely, considering the following minimization problem:
\begin{equation}
    \min \int_{\Omega}f(|\nabla\rho|)d\Omega,
\end{equation}
then the steepest descend equation may be computed to be 
\begin{equation}\label{steepdescend}
    \frac{\partial \rho}{\partial t} = \nabla\cdot(f'(|\nabla\rho|\frac{\nabla\rho}{|\nabla\rho|})).
\end{equation}
Obviously, (\ref{steepdescend}) is identical to (\ref{anisotropicdiffusion}) if 
\begin{equation}
    f'(x) = x\sigma(x).
\end{equation}
For example, the corresponding $f$ function of $\sigma$ function (\ref{sigma1}) is 
\begin{equation}
    f(x) = \frac{\sigma_0K^2}{2}\ln[1+(\frac{x}{K})^2]
\end{equation}

\subsection{rOMT with nonlinear diffusion}

In this section, we present our new rOMT formulation. We replace the diffusion in (\ref{ade}) by anisotropic diffusion in (\ref{anisotropicdiffusion}) and obtain the following formulation:
\begin{align}
    \inf_{\rho,v} &\int_0^1\int_{\Omega} \rho(t,x)|v(t,x)|^2dxdt+\beta\int_{\Omega}(\rho(1,x)-\rho_1(x))^2dx, \nonumber\\
    &\frac{\partial\rho}{\partial t}+\nabla\cdot(\rho v) = \nabla\cdot(\sigma(|\nabla\rho|)\nabla\rho),\label{anisotropicade} \\
    &\rho(0,x) = \rho_0(x). \nonumber
\end{align}

One may employ various versions of the $\sigma$ function and in this work, we choose the function given in (\ref{sigma1}). Note that, there are two parameters $\sigma_0$ and $K$ which may be tuned based on the data we use. 

Equation (\ref{anisotropicade}) may be written in conservation form as
\begin{equation*}
    \frac{\partial\rho}{\partial t}+\nabla\cdot(\rho( v-\sigma(|\nabla\rho|)\nabla\log\rho)) = 0,
\end{equation*}
and after defining an augmented velocity
\begin{equation*}
    v^{\mbox{aug}} = v-\sigma(|\nabla\rho|)\nabla\log\rho,
\end{equation*}
we derive a simple conservation form of equation (\ref{anisotropicade})
\begin{equation*}
    \frac{\partial\rho}{\partial t}+\nabla\cdot(\rho v^{\mbox{aug}}) = 0.
\end{equation*}

The Lagrangian representation $X = X(x,t)$ of the optimal trajectory for this rOMT with nonlinear diffusion model is given by 
\begin{equation}\label{lagrangian}
X(x,0) = x, \quad \frac{\partial X(x,t)}{\partial t} = v^{\mbox{aug}}_{\mbox{opt}}(X(x,t),t), 
\end{equation}
where 
\begin{equation}
    v^{\mbox{aug}}_{\mbox{opt}} = v_{\mbox{opt}}-\sigma(|\nabla\rho_{\mbox{opt}}|)\nabla\log\rho_{\mbox{opt}},
\end{equation}
and $v_{\mbox{opt}}$ and $\rho_{\mbox{opt}}$ denote the optimal solution of the rOMT with nonlinear diffusion model. 

In Section~\ref{results}, we exhibit the pathlines in Figure~\ref{C294pe1} and Figure~\ref{C294pe2} derived from the Lagrangian coordinates (\ref{lagrangian}).

\section{Numerical scheme}\label{numeric}
In this section, we focus on the numerical solution of the nonlinear diffusive rOMT model. The pipeline that comes from \cite{elkin2018glymphvis,koundal2020optimal} is based on the Gauss-Newton method:
\begin{enumerate}
    \item Give initial guess of $v$ at each time and spatial point.
    \item Use $v$, $\rho_0$ and the advection-diffusion equation (\ref{anisotropicade}) to calculate $\rho$ at each subsequent time step.
    \item Calculate the objective function (\ref{romtenergyfunc}), which we will denote with $\Gamma(v)$ as the discrete form.
    \item Calculate the gradient $g(v)$ and the Hessian matrix $H(v)$ of $\Gamma(v)$ with respect to $v$.
    \item Solve the descent direction $s$ by solving $H(v)s=-g(v)$.
    \item Do line search to find $l$ and update $v$ by setting $v=v+ls$.
    \item Repeat step 2-6 until the results attain the final condition.
\end{enumerate}

Space is discretized into a cell-center grid of size $n_x\times n_y\times n_z$ with a total number of $N$ cells, each with width $\Delta x$, height $\Delta y$ and depth $\Delta z$. Time is divided into $m$ intervals of length $\Delta t$ with $m+1$ time steps. Moreover, the superscript $0$ corresponds to initial time $t=0$, $M$ corresponds to final time $t=1$ and $dt\times m=1$. We use $\rho=[(\rho^0)^T,\dots,(\rho^m)^T]^T$ and $v=[(v^1)^T,\dots,(v^m)^T]^T$ to represent temporal density and velocity, respectively. Note that the velocity $v^i$ describes the velocity field from $(i-1)^{\mbox{th}}$ time step to $i^{\mbox{th}}$ time step.

\subsection{Advection-diffusion equation}
Here we describe the numerical scheme for equation (\ref{anisotropicade}).

The discrete form of equation (\ref{anisotropicade}) between time $t_n$ and $t_{n+1}$ is
\begin{equation}\label{dade}
    \frac{\rho^{n+1}-\rho^{n}}{\Delta t} + A(\rho,v) = D(\rho),
\end{equation}
where $A$ and $D$ are discretizations of advective and diffusive terms, respectively. We
will describe these in greater detail below. Following the work of Steklova and Haber \cite{steklova2017joint}, we split equation (\ref{dade}) into two parts,
\begin{align}
    &\frac{\rho^{\mbox{adv}}-\rho^{n}}{\Delta t} + A(\rho,v) = 0, \label{dae}\\
    &\frac{\rho^{n+1}-\rho^{\mbox{adv}}}{\Delta t} = D(\rho),\label{dde}
\end{align}
where $\rho^{\mbox{adv}}$ is an auxiliary variable. Simply by adding (\ref{dae}) and (\ref{dde}), we obtain the equation (\ref{dade}).
So far we have not chosen the time step of $\rho$ in the advective part $A(\rho,v)$ and diffusive part $D(\rho)$. We use a standard forward scheme, i.e. $\rho=\rho^n$ in our implementation.  Summarizing up to this point, to solve for the next time step density $\rho^{n+1}$, we first calculate $\rho^{\mbox{adv}}$ by solving equation ($\ref{dae}$) and use $\rho^{\mbox{adv}}$ and $\rho^n$ to calculate $\rho^{n+1}$ following equation ($\ref{dde}$).

For the advective part $A(\rho,v)$, we utilize a particle-in-cell method which is also how Steklova and Haber\cite{steklova2017joint} dealt with their advective part to solve equation (\ref{dae}):
\begin{equation}\label{rhoadv}
    \rho^{\mbox{adv}} = S(v)\rho.
\end{equation}
$S(v)$ is the averaging matrix with respect to $v$.

The basic idea of particle-in-cell method is moving density the $\rho_i$ in the cell center to the target $\rho_i^{\mbox{new}}$ according to its velocity $v_i$ and using its nearest neighbor cell centers to interpolate. 

The numerical techniques of solving equation (\ref{dde}) are based on hyperbolic
conservation laws and the theory of viscosity solutions \cite{sethian1989review,osher1988fronts,rudin1992nonlinear}, and we explicitly write $D$ in the next section.

\subsection{Anisotropic diffusion{}}
From now on, we explore in 3-dimension ($d=3$), following \cite{rudin1992nonlinear,you1996behavioral} and discretize the anisotropic diffusion as follows:
\begin{multline}
    D(\rho_{i,j,k}) = \Delta^x_-\{\sigma[\sqrt{(\Delta^x_+\rho_{i,j,k})^2+m^2(\Delta^y_+\rho_{i,j,k},\Delta^y_-\rho_{i,j,k})+m^2(\Delta^z_+\rho_{i,j,k},\Delta^z_-\rho_{i,j,k})}]\Delta^x_+\rho_{i,j,k}\}\\
    +\Delta^y_-\{\sigma[\sqrt{(\Delta^y_+\rho_{i,j,k})^2+m^2(\Delta^x_+\rho_{i,j,k},\Delta^x_-\rho_{i,j,k})+m^2(\Delta^z_+\rho_{i,j,k},\Delta^z_-\rho_{i,j,k})}]\Delta^y_+\rho_{i,j,k}\}\\
    +\Delta^z_-\{\sigma[\sqrt{(\Delta^z_+\rho_{i,j,k})^2+m^2(\Delta^x_+\rho_{i,j,k},\Delta^x_-\rho_{i,j,k})+m^2(\Delta^y_+\rho_{i,j,k},\Delta^y_-\rho_{i,j,k})}]\Delta^z_+\rho_{i,j,k}\}.
\end{multline}
Here, 
\begin{align*}
    &\Delta^x_-a_{i,j,k} = \frac{a_{i,j,k}-a_{i-1,j,k}}{\Delta x},\quad
    \Delta^x_+a_{i,j,k} = \frac{a_{i+1,j,k}-a_{i,j,k}}{\Delta x},\\
    &\Delta^y_-a_{i,j,k} = \frac{a_{i,j,k}-a_{i,j-1,k}}{\Delta y},\quad
    \Delta^y_+a_{i,j,k} = \frac{a_{i,j+1,k}-a_{i,j,k}}{\Delta y},\\
    &\Delta^z_-a_{i,j,k} = \frac{a_{i,j,k}-a_{i,j,k-1}}{\Delta z},\quad
    \Delta^z_+a_{i,j,k} = \frac{a_{i,j,k+1}-a_{i,j,k}}{\Delta z},\\
    &m(a,b) = \mbox{median}(a,b,0).
\end{align*}

We note that the solution of equation (\ref{dade}) may be written recursively:
\begin{equation}\label{sdade}
    \rho^{n+1} = S(v^n)\rho^n+\Delta tD(\rho^n).
\end{equation}

\subsection{Objective function $\Gamma(v)$}

A straightforward way \cite{elkin2018glymphvis,koundal2020optimal} to discretize the objective function $\Gamma(v)$ in (\ref{romtenergyfunc}) is 
\begin{equation}\label{denergyfunc}
    hd*\Delta t* \rho^T(I_m\otimes[I_N|I_N|I_N])(v\odot v)+\beta|\rho^m-\rho_T|^2.
\end{equation}
Here $hd=\Delta x*\Delta y*\Delta z$, $\rho,v$ are column vectors, $\otimes$ is Kronecker product and $\odot$ is Hadamard product.

\subsection{Gradient, hessian and sensitivity}
In order to apply the Gauss-Newton minimization procedure such as described in Steklova and Haber \cite{steklova2017joint}, we need expressions for the gradient $g(v)$ and the Hessian $H(v)$. Taking the gradient of (\ref{denergyfunc}) with respect to $v$, we find
\begin{equation}\label{gradient}
    g(v) = \frac{\partial\Gamma(v)}{\partial v} = hd*\Delta t*[2\rho^TM\diag(v)+(M(v\odot v))^TJ]+\beta(\rho^m-\rho_1)^T\frac{\partial\rho^m}{\partial v},
\end{equation}
where $M=I_m\otimes[I_N|I_N|I_N]$, matrix $J=(J^k_{j})$. Here $J^k_j=\frac{\partial\rho^k}{\partial v^j}$, $k=1,\dots,m$ and $j=0,\dots,m-1$. 

The Hessian matrix is 
\begin{multline}
        H(v) = \frac{\partial g}{\partial v} = hd*\Delta t*[2\rho^T\nabla(M\diag(v))+2\nabla(\rho)M\diag(v)+M(v\odot v)\nabla J+\nabla[M(v\odot v)]J]\\
        +\beta[(\frac{\partial\rho^m}{\partial v})^T(\frac{\partial\rho^m}{\partial v})+(\rho^m-\rho_1)\frac{\partial^2\rho^m}{\partial v^2}].
\end{multline}
Numerically we approximate the Hessian by 
\begin{align}\label{hessian}
    H(v) & = 2hd*\Delta t*\rho^T\nabla(M\diag(v))+\beta(\frac{\partial\rho^m}{\partial v})^T(\frac{\partial\rho^m}{\partial v})\nonumber\\
    & = 2hd*\Delta t*\diag(\rho^TM) + \beta(\frac{\partial\rho^m}{\partial v})^T(\frac{\partial\rho^m}{\partial v}).
\end{align}

In the formulae for the gradient (\ref{gradient}) and Hessian (\ref{hessian}), we still need to know the sensitivity of the density $\rho$ with respect to the velocity $v$. 
We recall equation (\ref{sdade})
\begin{equation*}
    \rho^{n+1} = S(v^n)\rho^n+\Delta tD(\rho^n).
\end{equation*}
From that, the sensitivity can be calculated as below:
\begin{equation}\label{sensitivity}
\frac{\partial\rho^k}{\partial v^j} = \left\{
        \begin{array}{ll}
            S(v^{k-1})\frac{\partial\rho^{k-1}}{\partial v^j}+\Delta t D'(\rho^{k-1})\frac{\partial\rho^{k-1}}{\partial v^j} & \quad k \ge j+2 \\
            \frac{\partial}{\partial v^j}(S(v^j)\rho^j) & \quad k = j+1\\
            0 & \quad k\le j
        \end{array}
    \right.
\end{equation}

\section{Experimental results}\label{results}
In this section, we test our proposed methodology on 3D DCE-MRI data derived from \cite{chen2022cerebral}. In this dataset, rats were anesthetized, and a Gd-tagged tracer was injected into the cerebrospinal fluid (CSF). The rat underwent dynamic 3D MRI scanning  every 5 minutes to collect a total 29 3D brain images with a voxel size of 100$\times$106$\times$100. Post-processing of the DCE-MRI data included head motion correction, intensity normalization, and voxel-by-voxel conversion to percentage of baseline signal. In our experiment, we chose a 12-month-old wild type rat for demonstrating the results. 

The new algorithm was run for data covering a 100-minute time period (60 minutes to 160 minutes) which includes 23 frames, and we used every other image as inputs to reduce runtime, leaving 12 frames for the numerical experiment. We use $I_n, n=1,\dots,12$ to represent these frames. To derive the interpolations, we applied our model between each of two consecutive frames, i.e. $I_k$ and $I_{k+1}$. To ensure continuity, (except for the first step), the initial density originates from the previous step. For example, if we are considering the problem between $I_2$ and $I_3$, and we will use the final density $I_2'$ calculated between $I_1$ and $I_2$ as the new initial density here and apply our model between $I_2'$ and $I_3$. One of the metrics that can measure the model accuracy is the error between the final density $I_k'$ and the ground truth $I_k$ at each step. 

\begin{figure}
    \centering
    \includegraphics[width=0.32\linewidth]{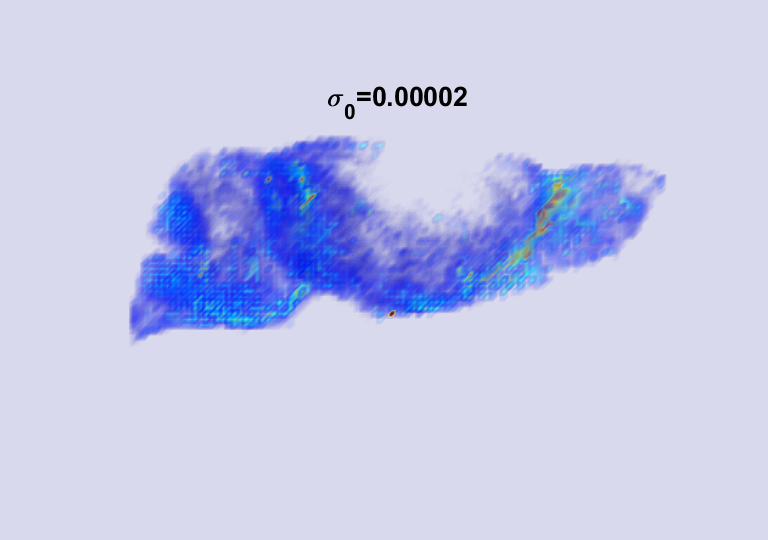}
    \includegraphics[width=0.32\linewidth]{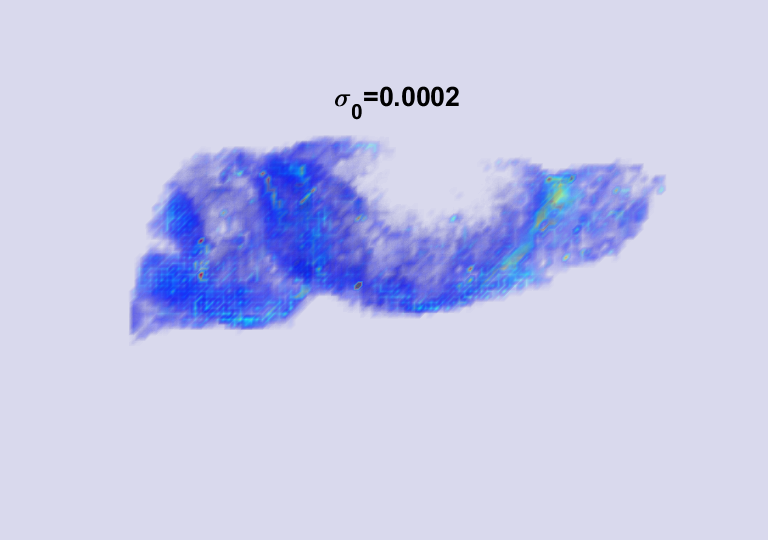}
    \includegraphics[width=0.32\linewidth]{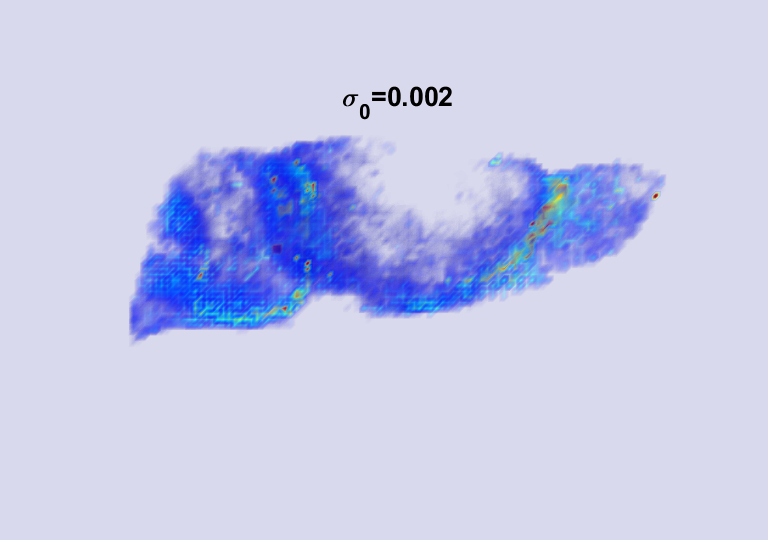}
    \includegraphics[width=0.32\linewidth]{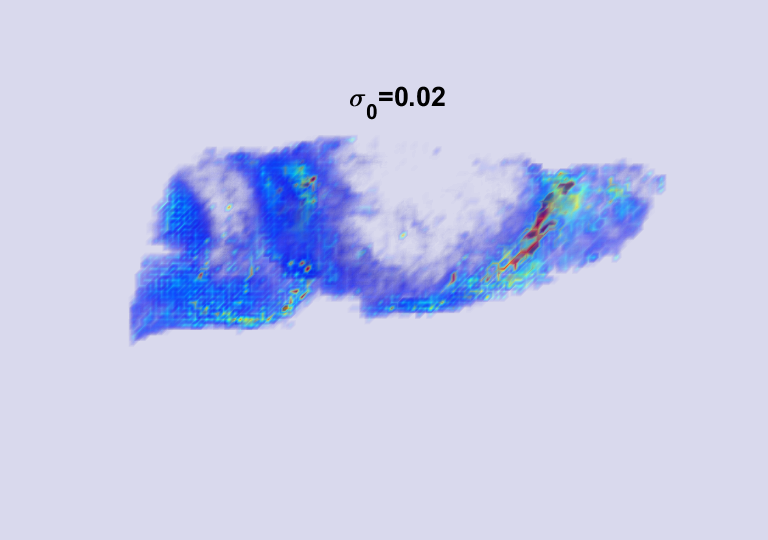}
    \includegraphics[width=0.32\linewidth]{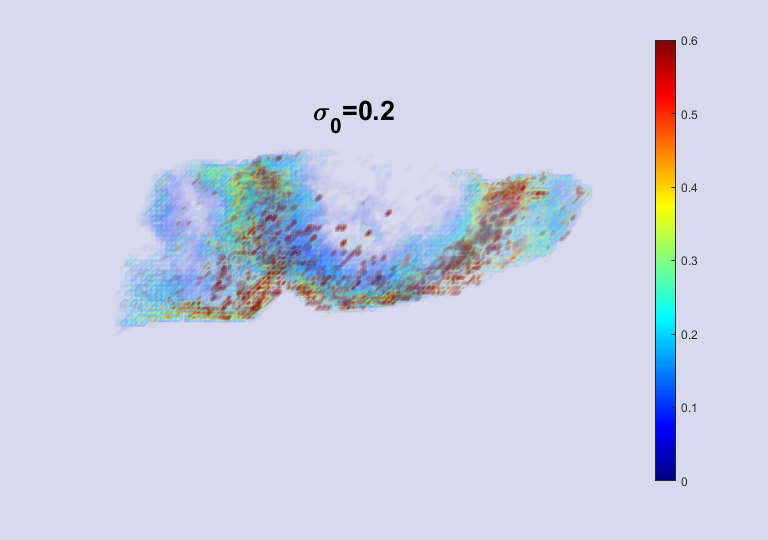}
    \caption{Speed map for different $\sigma_0$'s. The maximal limit of the color bar is $0.6$. The first three speed maps exhibit a stable trend. The last two speed maps with higher values of diffusion dramatically (and erroneously) increase speed suggesting that $\sigma_0$ is too large.}
    \label{speedmap}
\end{figure}

\begin{figure}
    \centering
    \includegraphics[width=0.8\linewidth]{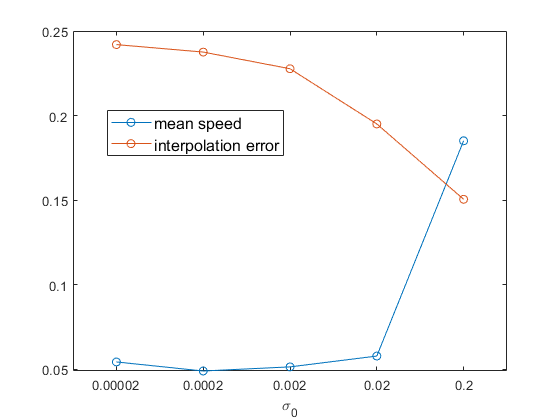}
    \caption{Mean speed (blue line) and interpolation error (orange line) of different $\sigma_0$'s. The interpolation error is the relative error between interpolated frames and data image of the last frame. The interpolation error reflects the closeness between interpolations from rOMT and the data image. Lower interpolation error means more accurate the rOMT is fitting the real data. This figure shows larger $\sigma_0$ has better interpolation error but when $\sigma_0$ goes to $0.2$, the mean speed accelerates dramatically, which is unrealistic given previous data of the expected magnitude of solute transport in brain tissue.}
    \label{msie}
\end{figure}

Here we are using $\sigma$ function ($\ref{sigma1}$) with $\sigma_0=0.002$. The choice of $\sigma_0$ follows \cite{chen2022cerebral}. We tested rOMT on the 3D DCE-MRI data set with $\sigma_0=0.00002,0.0002,0.002,0.02,0.2$. The speed maps in Figure \ref{speedmap} show a stable trend between $\sigma_0=0.00002$ and $\sigma_0=0.002$ and among these three $\sigma_0$ ($0.00002$,$0.0002$ and $0.002$), $0.002$ has the minimal interpolation error (see Figure~\ref{msie}). 

\begin{figure}
    \centering
    \includegraphics[width=0.98\linewidth]{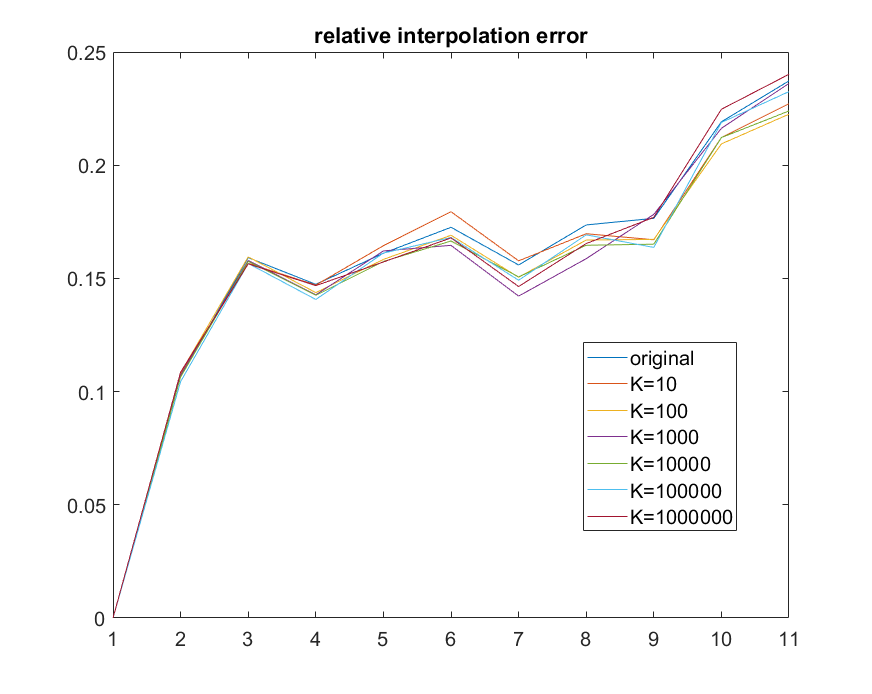}
    \caption{Relative interpolation error plot for different parameter $K$'s. Original means constant diffusion coefficient, i.e. $K=\infty$.}
    \label{error1}
\end{figure}

\begin{figure}
    \centering
      \foreach \t in {1,2,3,4,5,6}{
      \includegraphics[width=0.48\linewidth]{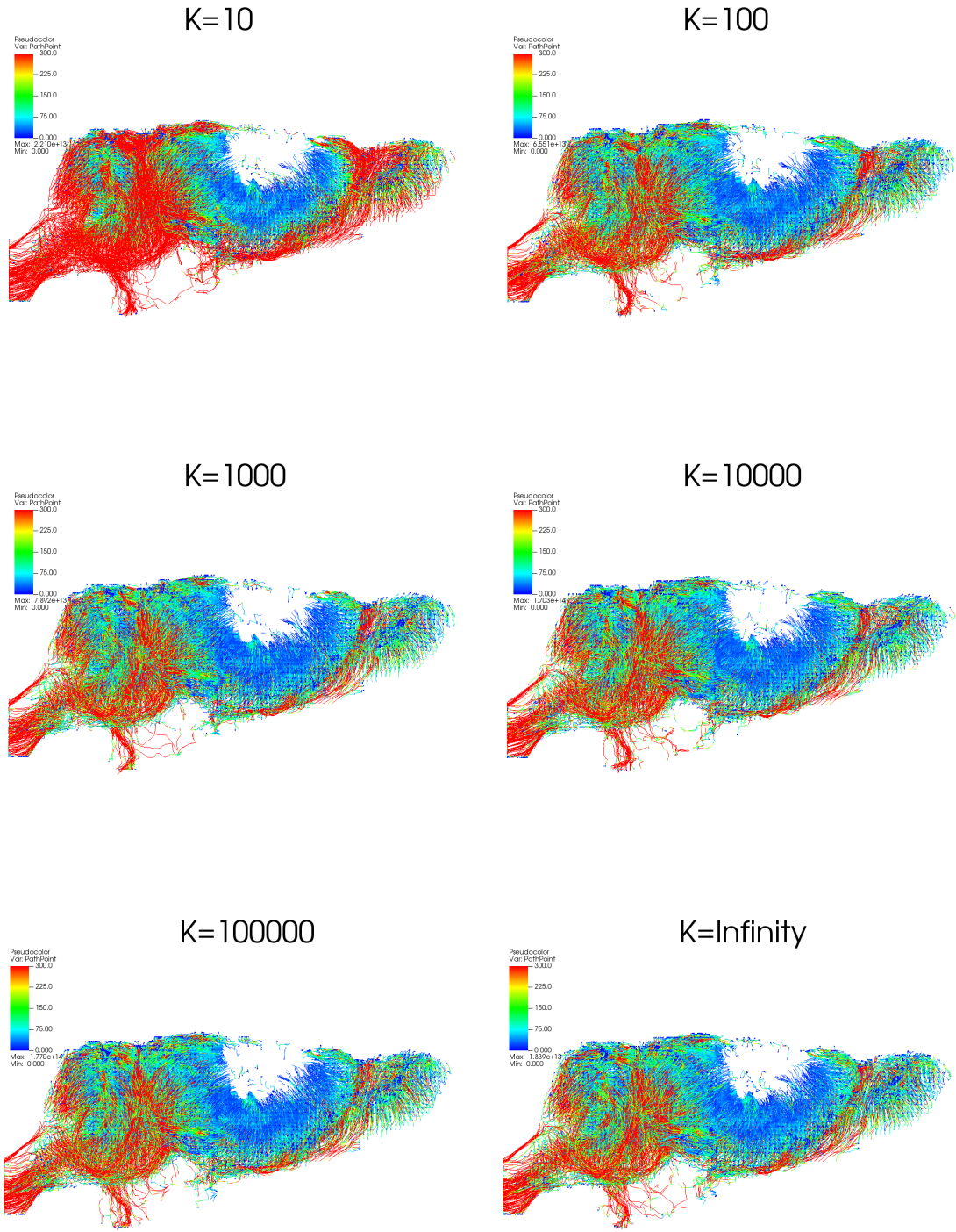}
      }
    \caption{Pathlines endowed with P\'{e}clet Number shown in the lateral view plane. Parameter $K=10,100,1000,10000,100000,\infty$. The maximal limit of color bar is $300$. When $K$ is small, the advective (red) pathline dominates in CSF rich areas.}
    \label{C294pe1}
\end{figure}

\begin{figure}
    \centering
      \foreach \t in {1,2,3,4,5,6}{
      \includegraphics[width=0.48\linewidth]{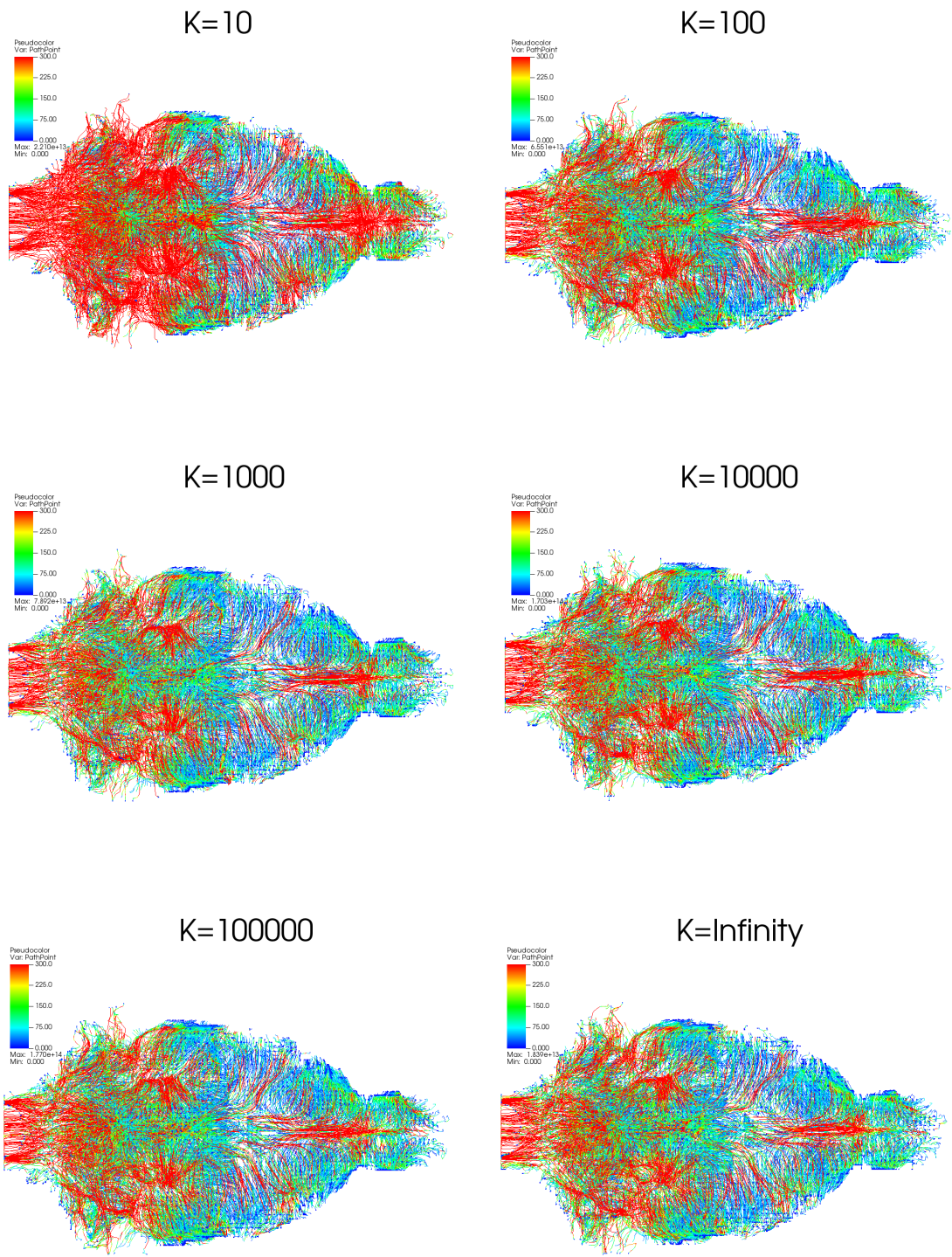}
      }
    \caption{P\'{e}clet number endowed pathlines shown in ventral view plane. Parameter $K=10,100,1000,10000,100000,\infty$. The maximal limit of the color bar is $300$.}
    \label{C294pe2}
\end{figure}

We computed pathlines based on Lagrangian coordinates (\ref{lagrangian}). We compared different $K$'s and the results are shown in Figures~\ref{error1}-\ref{C294pe2}. Figure~\ref{error1} shows the relative error
\begin{equation*}
    e = \frac{|I'-I|_2}{|I|_2}
\end{equation*}
on each frame with different $K$'s. The $x$-axis represents the indices of frames and the $y$-axis is the relative error. From Figure~\ref{error1}, we observe that rOMT with anisotropic diffusion has similar accuracy as the original rOMT model. Figure~\ref{C294pe1} compares the P\'{e}clet number along pathlines in the right lateral view plane for different $K$'s. Further, Figure~\ref{C294pe2} shows the ventral surface of the brain. Red color represents larger P\'{e}clet numbers (advection dominant) and blue represents smaller P\'{e}clet numbers (diffusion dominant). As shown in Figure~\ref{C294pe1} and Figure~\ref{C294pe2}, a smaller $K$ value results in more advection dominated transport in `surface' areas of the brain which corresponds to the CSF compartment. When we set $K=\infty$, then clearly $\sigma(x)=\sigma_0$, since 
\begin{equation*}
    \lim_{K\to\infty}\sigma_0\frac{1}{1+(\frac{x}{K})^2}=\sigma_0.
\end{equation*}


\section{Discussion}\label{discussion}
In this paper, we proposed a novel extension of the rOMT model. Specifically, we replaced the linear diffusion term in the advection-diffusion equation by a nonlinear diffusion term based on the Perona-Malik anisotropic diffusion approach. The updated model was tested on glymphatic DCE-MRI data comparing different parameter $K$'s in the conductivity coefficient ($\sigma$) function and we observed that smaller $K$ yields increased number of advective pathlines in CSF rich areas. More uniform advective solutes flow in the CSF compartment including at the level of the basal cisterns, ambient cistern and subarachnoid space above the cerebellum may be more biologically realistic.

This paper only applied the model on glymphatic DCE-MRI data, but it can be generally applied to other types of biological imaging data. In the future, we plan to apply our approach to tumor vasculature imagery also derived from DCE-MRI, since the mass (tracer) is injected and may leak, we also plan to explore an unbalanced version of rOMT with nonlinear diffusion.

\subsubsection*{Acknowledgments}
This research was funded in part by AFOSR grant FA9550-20-1-0029, NIH grant R01-AG048769, a grant from Breast Cancer Research Foundation BCRF-17-193, Army Research Office grant  W911NF2210292, and a grant from the Cure Alzheimer's Foundation.

\bibliographystyle{plain}
\bibliography{refs}

\end{document}